\documentclass{frontiersSCNSflf} 

\usepackage{url,hyperref,lineno,microtype}
\usepackage[onehalfspacing]{setspace}

\def\keyFont{\fontsize{8}{11}\helveticabold }
\def\firstAuthorLast{La Franca {et~al.}} 
\def\Authors{Fabio La Franca\,$^{1,*}$, Francesca Onori\,$^{2}$, Federica Ricci\,$^1$,  Stefano Bianchi\,$^1$, Alessandro Marconi\,$^3$, Eleonora Sani\,$^4$, Cristian Vignali\,$^5$}
\def\Address{$^{1}$Dipartimento di Matematica e Fisica, Universit\`a degli Studi Roma Tre, via della Vasca Navale 84, 00146, Roma, Italy\\
$^{2}$ Netherlands Institute for Space Research, SRON, Sorbonnelaan 2, NL 3584 CA Utrecht, the Netherlands\\
$^{3}$ Dipartimento di Fisica e Astronomia, Universit\`a degli Studi di Firenze, Via G. Sansone 1, 50019, Sesto Fiorentino, Italy\\
$^{4}$ European Southern Observatory, Alonso de Cordova 3107, Casilla 19, Santiago 19001, Chile\\
$^{5}$ Dipartimento di Fisica e Astronomia, Universit\`a di Bologna, viale Berti Pichat 6/2, 40127, Bologna, Italy
}
\def\corrAuthor{Fabio La Franca}
\def\corrAddress{Dipartimento di Matematica e Fisica, Universit\`a degli Studi Roma Tre, via della Vasca Navale 84, 00146, Roma, Italy}
\def\corrEmail{lafranca@fis.uniroma3.it}

\begin{document}
\onecolumn
\firstpage{1}

\title[Faint BLR Detection in NGC 6221]{Detection of Faint BLR Components in the Starburst/Seyfert  Galaxy  NGC 6221 and Measure of the Central BH Mass} 

\author[\firstAuthorLast ]{\Authors} 
\address{} 
\correspondance{} 

\extraAuth{}

\maketitle


\begin{abstract}
 In the last decade, using single epoch virial based techniques in the optical band, it has been possible to measure the central black hole mass on large type 1 Active Galactic Nuclei (AGN) samples. However these measurements use the width of the
 broad line region as a proxy of the virial velocities and  are therefore difficult to be carried out on those obscured (type 2) or low luminosity AGN where the nuclear component does not dominate in the optical.  Here we present the optical and near infrared spectrum of the starburst/Seyfert  galaxy NGC 6221, observed with X-shooter/VLT. Previous observations of 
 NGC 6221 in the X-ray band show an absorbed (${\rm N_H=8.5 \pm 0.4 \times 10^{21} cm^{-2}}$)  spectrum typical of a type 2 AGN with 
luminosity log(L$_{14-195}/ {\rm erg~s^{-1}}$) = 42.05, while in the optical band its spectrum is typical of a reddened (${\rm A_V=3}$) starburst.
 Our deep X-shooter/VLT observations have allowed us to detect
 faint broad emission in the H$\alpha$, HeI and Pa$\beta$ lines (FWHM $\sim$1400-2300 km s$^{-1}$) confirming previous studies indicating that  
 NGC 6221 is a reddened starbust galaxy which hosts an AGN.
 We use the measure of the broad components to provide a first estimate of its central black hole mass
 (${\rm M}_{\rm BH} = 10^{6.6\pm0.3} {\rm M}_\odot$, $\lambda_{\rm Edd}=0.01-0.03$), obtained using recently calibrated virial relations suitable for moderately  obscured (N$_{\rm H}<10^{24}~{\rm cm}^{-2}$) AGN.

 

\tiny
 \keyFont{ \section{Keywords:} extragalactic astronomy, galaxies, active galactic nuclei, black holes} 
\end{abstract}

\section{Introduction}

%
%
Nowadays there is robust evidence that  every galaxy hosts a supermassive black hole (SMBH; M$_{\rm BH}$=10$^6$-10$^9$ M$_{\odot}$) whose mass scales with the hosting galaxy bulge properties \citep[mass, luminosity and stellar dispersion; e.g.][]{Ferrarese00, Gebhardt00, Marconi03, Sani11}. The existence of these scaling relationships implies that the evolution of the galaxy and the growth of SMBHs are intricately tied toghether (AGN/galaxy co-evolution scenario). 
In order to obtain a clear picture of the AGN/galaxy co-evolution, it is important to accurately derive the shape and the evolution of both the AGN luminosity function and the SMBH mass function  in a consistent way. While the complete AGN luminosity function is fairly well measured up to $z\sim$4, this is not the case for the SMBH mass function. 

Recently it has been possible to obtain some estimates of the SMBH mass function for large samples of type 1 AGN (AGN1). In this class of AGN the broad line region (BLR) is visible in the rest-frame optical band and this allows the use of virial methods to derive in a direct way the AGN BH mass (\cite{greene07b}, \cite{Kelly09}, \cite{merloni10}, \cite{bongiorno14}). However this kind of measurements cannot be applied on narrow line type 2 AGN (AGN2), where the BLR is not visible in the optical spectrum because of dust absorption.
Moreover, nowadays there is growing evidence that AGN1 and AGN2 are intrinsically different populations \citep[see e.g.][]{elitzur12},  having, on average, different luminosities
\citep[smaller for AGN2;][]{lawrence82,ueda03,lafranca05}, different accretion rates \citep[smaller for AGN2;][]{Winter10},  different Eddington ratios \citep{lusso12}, different clustering, halo mass properties and merger rates \citep{allevato14,lanzuisi15a}.
It is, therefore,  very important to find a method to easily and reliably estimate the BH mass of the AGN2. Recently \citet{lafranca15} have calibrated a  new virial relationship suitable for moderately absorbed/obscured AGN2, which is based on the measure of the Full Width at Half Maximum (FWHM) of the BLR component of the Pa$\beta$ emission line (emitted in the near infrared; NIR) together with the measure of the hard (14-195 keV) luminosity. We here present the X-shooter/VLT UV-Optical-NIR spectrum of the starburst/Seyfert 2 galaxy NGC 6221.
We fitted the regions of the H$\beta$+[OIII], H$\alpha$+[SII], HeI and Pa$\beta$ lines in order to try to
detect faint BLR components and eventually estimate the BH mass. Troughout the paper uncertainties are given at the 1$\sigma$ confidence level.
We adopt a $\Omega_m=0.3$, $\Omega_\lambda=0.7$, ${\rm H_0 = 70~ km~ s^{-1}~ Mpc^{-1}}$ cosmology.

\section{The galaxy NGC 6221}
{

NCG 6221  ($\alpha$= 16$^h$ 52$^m$ 46.3$^s$, $\delta$ = -59$^\circ$ 13$^m$ 07$^s$; J2000)  is a nearby
\citep[$z$=0.0050;][]{koribalski04} spiral galaxy (${\rm 22\times 15}$ kpc$^2$) classified as Sbc by \citet{dressler78}, Sbc(a) by \citet{sandage81} and as SBc(s), with position angle (PA) 5$^\circ$  and inclination 44$^\circ$, by \citet{devaucouleurs91}. The bar, which is clearly visible in the optical and infrared images \citep{laustsen87,sandage94}, lies at a PA of 118$^\circ$ \citep{pence84} and has  a length of $\sim$6 kpc. Large amount of dust is visible in both spiral arms as well as along the bar.
NGC 6221 forms an apparent physical pair with the late type spiral NGC 6215 wich is about 110 kpc distant and is also possibly interacting with two low-surface brightness galaxies nearby \citep{koribalski04}. The rotation curve reaches a maximum velocity of 160 km/s at $\sim$10 kpc of radius, corresponding to an enclosed mass of 6$\times$10$^{10}$ M$_\odot$ \citep[8$\times$10$^{10}$ M$_\odot$ at 20 kpc;][]{koribalski04}.

NGC 6221 is an example of the so called X-ray Bright Optical Normal Galaxies \citep[XBONG;][]{fiore00,civano07,moran96}.
This classification comes from the comparison between its optical spectrum, which is not AGN-like (starburst in this case), and its X-ray data, where the AGN is revealed \citep{levenson01}.
In the X-ray band NGC 6221 shows an absorbed (${\rm N_H=8.5 \pm 0.4 \times 10^{21} cm^{-2}}$) variable (on timescales of days and years) spectrum typical of type 2 AGN with a 2-10 keV intrinsic luminosity  ${\rm L_{2-10} = 6.6\times 10^{41}~erg~s^{-1}}$ \citep[][and Bianchi et al. in prep]{levenson01}. NGC 6221 has been observed by SWIFT/BAT in the 14-195 keV band and a luminosity log(L$_{14-195}/ {\rm erg~s^{-1}}$) = 42.05 was measured \citep[70-month catalogue;][]{Baumgartner13}.

The net nuclear spectrum of NGC 6221 is typical of a reddened (${\rm A_V=3}$) starburst \citep{phillips79,morris88,storchi95}.
A possible sign of nonstellar activity at these wavelenghts is an $[$OIII$]$ component broader than, and blueshifted with respect to, H$\beta$. This feature, as well as the early detection of NGC 6221 as an X-ray source \citep{marshall79} motivated \citep{veron81} to propose a composite Seyfert 2/starburst scenario \citep[see also][]{pence84,boisson86,dottori96,levenson01}.  The radio continuum emission consists of a bright nucleus and diffuse circumnuclear emission extending as far as the optical disc, resulting to a star formation rate of $\sim$15 M$_\odot$ yr $^{-1}$ \citep{koribalski04}. 
\citet{cidfernandes03} give, indeed, a (fairly short) mean starburst age of 10$^{7.4\pm 1.1}$ yr for the central 10$\times$20 arcsec$^2$ of NGC 6221.
 
Kinematical studies of NGC 6221, using several long-slit spectra of the H$\alpha$ line emission out to radii of 80 arcsec \citep{pence84, vegabeltran98}, revealed non-circular motion of ionized gas, possibly as a result of streaming motions along the bar and tidal interaction with NGC 6215. 
The isovelocity contour map shows a conspicuous S-shaped pattern of closely spaced contours near the minor axis indicating a very sudden change in velocity of about 150 km s$^{-1}$ at the position of the dust lanes in the bar. The gas is moving radially outward as it approaches the dust lane, and then inward after passing trough the shock. This transition occurs in a narrow region, less than 200 pc wide \citep[see the cartoon of the model described in Fig. 5 of][]{pence84}. 

In the optical spectrum, broad and blue components are observed which  dominates the $[$OIII$]$ profile (but not the H$\beta$ line) extending to more than 1000 km s$^{-1}$. The $[$OIII$]$ to H$\beta$ ratio vary
from 0.2 at the line peak (typical of normal HII regions) to more than 3 in the blue wing. Since the $[$OIII$]$ to H$\beta$ ratio is an indicator of the excitation conditions, either from shocks or highly ionizing radiation
\citep{baldwin81}, the gas with highest blueshift shows the highest excitation. Moreover the $[$OIII$]$ profile changes as a function of the distance from the nucleus. In the nuclear (1.5$''\times1.5''$; 1 arcsec $\sim$ 103 pc at the distance of NGC 6221) spectrum, the broad and blue component dominates. 
The fitting of the $[$OIII$]$ profile  yelds a broad component with FWHM$\sim$600 km/s shifted by approximately 230 km/s with respect to the narrow
unresolved component, while in the H$\beta$ the narrow component dominates. At variance, at distance of 500 pc from the nucleus the broad and blue wing of the H$\beta$ line is no longer visible, and both the $[$OIII$]$ and H$\beta$ profiles
looks narrow and very similar. However it should be noticed that also in both the  H$\alpha$ and H$\beta$ profiles a weak broad component of FWHM$\sim$ 500-600 km/s, blueshifted by 150-250 km/s 
with respect to an unresolved narrower component (whose intrinsic FWHM$\sim$ 100 km/s) was observed \citep{levenson01}.
As reported by \citet{levenson01} all the total nuclear emission-line ratios are consistent with a reddened (${\rm A_V=3}$; ${\rm A_V=2.5}$ once corrected for the Milky Way abpsorption) starburst classification on conventional diagnostic diagrams \citep{veilleux87}. Similar values of the extinction have been measured by \citet{ramosalmeida11} by spectral energy distribution (SED) fitting in the NIR.
No indications of broad (FWHM$>$ 1000 km s$^{-1}$) permitted emission lines were found both in the optical  and in the NIR \citep[][and references therein]{levenson01}.

All these observations motivated \citet{levenson01} to draw a scenario (see the cartoon in their Fig. 6) where a face on type 1 AGN, surrounded by a
starburst galaxy (responsible for the reddening in the optical and for the absorption in X-ray band),  is observed.
So that, if it were possible to turn off the starburst, NGC 6221 would be classified as a Seyfert 1 galaxy \citep[as previously proposed by][]{fabian98}. 

As under the above hypothesis, or any other model in which NGC 6221 hosts a reddened AGN, it could be possible to detect the faint BLR components  with deep NIR spectroscopy, we have decided to observe the nucleus of NGC 6221 with X-shooter at the VLT.

}
\section{Data and spectral line analysis}

\begin{table}[h]
\textbf{\refstepcounter{table}\label{tab:1} Table \arabic{table}.}{
Emission line fitting parameters}

\processtable{ }
{\begin{tabular}{llcccc}\toprule
Line & Comp. & EW & FWHM & $\Delta$V \\
              & & (\AA) & (km s$^{-1}$) & (km s$^{-1}$) \\\midrule
H$\beta$ & N& \phantom{8}4.7$\pm$0.2  &\phantom{88}79$^{+1\phantom{88}}_{-1}$&-\\
H$\beta$ & I & 12.9$\pm$0.2& \phantom{8}146$^{+1\phantom{88}}_{-4}$&\phantom{8}31$\pm$2\phantom{8}\\
H$\beta$ & I&  \phantom{8}7.8$\pm$0.2& \phantom{8}541$^{+13\phantom{8}}_{-13}$&195$\pm$1\phantom{8}\\
\\
$[$OIII$]$4958.9 & N &\phantom{8}2.1$\pm$0.5&  \phantom{8}177$^{+4\phantom{88}}_{-2}$  &-\\
$[$OIII$]$4958.9 & I$^a$ &  \phantom{8}0.6$\pm$0.2  & \phantom{8}125$^{+5\phantom{88}}_{-5}$&166$\pm$1\phantom{8}\\
$[$OIII$]$4958.9 & I & \phantom{8}1.4$\pm$0.4 &   \phantom{8}212$^{+3\phantom{88}}_{-4}$&346$\pm$1\phantom{8}\\
$[$OIII$]$4958.9 & I &\phantom{8}3.0$\pm$0.7 &  \phantom{8}{851}$^{+16\phantom{8}}_{-10}$ & 261$\pm$11\\
\\
$[$OIII$]$5006.8 & N & \phantom{8}6.4$\pm$1.6& \phantom{8}177$^{+4\phantom{88}}_{-2}$ & - \\
$[$OIII$]$5006.8 & I$^a$ &\phantom{8}2.0$\pm$0.5  & \phantom{8}125$^{+5\phantom{88}}_{-5}$ & 166$\pm$1\phantom{8}\\
$[$OIII$]$5006.8 & I & \phantom{8}4.3$\pm$1.1 & \phantom{8}212$^{+3\phantom{88}}_{-4}$ & 346$\pm$1\phantom{8}\\
$[$OIII$]$5006.8 & I & \phantom{8}9.7$\pm$2.4 & \phantom{8}{851}$^{+16\phantom{8}}_{-10}$ & 261$\pm$11\\
\midrule
$[$NII$]$6548.0 & N & \phantom{8}9.9$\pm$0.1 & \phantom{88}90$^{+4\phantom{88}}_{-3}$  & -\\
$[$NII$]$6548.0 & I &\phantom{8}8.8$\pm$0.2 & \phantom{8}192$^{+8\phantom{88}}_{-13}$  & \phantom{8}21$\pm$1\phantom{8} \\
$[$NII$]$6548.0 & I & 11.3$\pm$0.2& \phantom{8}500$^{+12\phantom{8}}_{-7}$    &172$\pm$2\phantom{8}\\
$[$NII$]$6548.0 & I$^b$ & \phantom{8}0.4$\pm$0.1 &  \phantom{88}77$^{+16\phantom{8}}_{-8}$   &375$\pm$5\phantom{8}   \\
\\
H$\alpha$ & N & 48.0$\pm$0.2 &  \phantom{88}90$^{+4\phantom{88}}_{-3}$  & -  \\
H$\alpha$ & I &  50.9$\pm$0.3& \phantom{8}192$^{+8}_{-13\phantom{8}}$  & \phantom{8}21$\pm$1\phantom{8}\\
H$\alpha$ & I &45.7$\pm$0.3 & \phantom{8}500$^{+12}_{-7\phantom{88}}$ &172$\pm$2\phantom{8}  \\
H$\alpha$ & I$^b$ &  \phantom{8}1.1$\pm$0.1& \phantom{88}77$^{+16\phantom{8}}_{-8}$ &375$\pm$5\phantom{8} \\
H$\alpha$ & B & 59.0$\pm$0.5& 1630$^{+12\phantom{8}}_{-11}$ &- \\
\\
$[$NII$]$6583.4 & N &29.9$\pm$0.2& \phantom{88}90$^{+4\phantom{88}}_{-3}$ & -  \\
$[$NII$]$6583.4 & I & 26.6$\pm$0.2&  \phantom{8}192$^{+8}_{-13\phantom{8}}$& \phantom{8}21$\pm$1\phantom{8} \\
$[$NII$]$6583.4 & I & 33.9$\pm$0.2&  \phantom{8}500$^{+12\phantom{8}}_{-7}$& 172$\pm$1\phantom{8}\\
$[$NII$]$6583.4 & I$^b$ & \phantom{8}1.1$\pm$0.1& \phantom{88}77$^{+16\phantom{8}}_{-8}$& 375$\pm$1\phantom{8} \\
\midrule
HeI 10830.2& N & 10.7$\pm$0.9 & \phantom{8}141$^{+1\phantom{88}}_{-1}$      &   - \\
HeI 10830.2& I & 21.7$\pm$1.8& \phantom{8}343$^{+9\phantom{88}}_{-9}$      & \phantom{8}52$\pm$4\phantom{8}\\
HeI 10830.2& B & 29.6$\pm$2.5 &  2142$^{+110}_{-141}$ &  -\\
\\
Pa$\gamma$ & N &  \phantom{8}8.6$\pm$0.7& \phantom{8}141$^{+1\phantom{88}}_{-1}$        &-  \\
Pa$\gamma$ & I &  \phantom{8}2.0$\pm$0.4 & \phantom{8}343$^{+9\phantom{88}}_{-9}$       &213$\pm$27 \\
Pa$\gamma$ & B &  \phantom{8}9.2$\pm$1.0& 1433$^{+70\phantom{8}}_{-70}$ & -   \\
\midrule
$[$FeII$]$12566.8& N &  \phantom{8}5.9$\pm$0.2& \phantom{8}141$^{+1\phantom{88}}_{-1}$      &  -   \\
$[$FeII$]$12566.8& I & 10.3$\pm$0.3& \phantom{8}483$^{+12\phantom{8}}_{-12}$ & 176$\pm$5\phantom{8} \\
\\
Pa$\beta$ & N & 18.6$\pm$0.2 & \phantom{8}141$^{+1\phantom{88}}_{-1}$     & -  \\
Pa$\beta$ & I &  \phantom{8}9.5$\pm$0.3   & \phantom{8}483$^{+12\phantom{8}}_{-12}$ &176$\pm$5\phantom{8} \\
Pa$\beta$ & B &  20.5$\pm$0.8&  2257$^{+99\phantom{8}}_{-93 }$&-   \\

\botrule
\end{tabular}}{}
$^{a,b}$: these lines, although narrower 
than their narrow component, are classified as intermediate as they show
a significant blueshift with respect   to the systemic redshift of the galaxy.

\end{table}

NGC 6221 was observed on April 24th 2014 with X-shooter at the VLT. X-shooter is a single target spectrograph covering in a single exposure the spectral range from the UV to the K band (300$-$2500 nm) \citep{Vernet11}. The instrument operates at intermediate resolutions, R=4000$-$18000, depending on the wavelength and the slit width.  Ten images with exposures, in the NIR band, of $\sim$290 s each, were acquired.
A 1.0$''$$\times$11$''$  slit for the ultraviolet and blue (UVB) arm and a 0.9$''$$\times$11$''$ slit for the optical (VIS) and NIR arms were used,  corresponding to a spectral resolution R=$\Delta \lambda/\lambda$=4350 for the UVB arm,  R=$\Delta \lambda/\lambda$=7450 for the VIS arm and R=$\Delta \lambda/\lambda$=5300 for the NIR arm, and to a velocity uncertainty  of $\Delta v$$\sim$70/40/60 km/s at redshift 0,
in the UVB/VIS/NIR arms, respectively. The data reduction has been carried out using the REFLEX X-shooter pipeline \citep{freudling13}.

We have analysed those emission lines where faint BLR components could be present: i.e. H$\beta$, H$\alpha$, HeI, Pa$\gamma$ and Pa$\beta$ lines. The air rest-frame wavelengths were used \citep{morton91}.
The 1$\sigma$ uncertainties provided by the data reduction pipelines, and compared with  featureless regions of the spectra, were used to carry out the line fitting using XSPEC 12.7.1 \citep{arnaud96}. At 5100 \AA\  the spectrum has a S/N$\sim$55 per resolution element.
The local
continuum was always modelled with a power-law and subtracted, then all significant (using the F-test) components were modelled with Gaussian profiles. All measurements were performed in the redshift corrected spectrum (i.e.
in the object rest frame). We have identified as narrow (N) all the components having widths less than $\sim$200 km s$^{-1}$, well centered with
the line profiles, and usually compatible with the forbidden lines widths (i.e. belonging to the narrow line region, NLR). At variance, the widest components of the permitted HI and HeI lines, significantly larger than the narrow components, have been classified as broad (B) and then associated to the BLR. Other
intermediate (I) width components \citep[all blueshifted with respect to the narrow components. See previous results by ][]{levenson01} were also identified .
The narrow components of permitted lines (HI and HeI) have been modelled by imposing the same FWHM (in the velocy space) found for the narrow components of the forbidden lines
measured  in the same spectral band. In the optical we have imposed that the intensity ratios between the [OIII]4959 and [OIII]5007 and between [NII]6548 and [NII]6583 satisfied the expected 1:2.99 relation \citep{osterbrock06} and that the central wavelenghts of all lines shared the same systemic redshift. When intermediate components were found, we have fixed their FWHM and their blueshift to be equal to those found for the corresponding intermediate component of the most intense forbidden line in the same spectral region. The main fitting parameters (equivalent width, EW, FWHM and the velocy off-set with respect to the NLR rest frame, $\Delta$V) are listed in  Table \ref{tab:1}.

\begin{figure}
\begin{center}
\includegraphics[width=17cm]{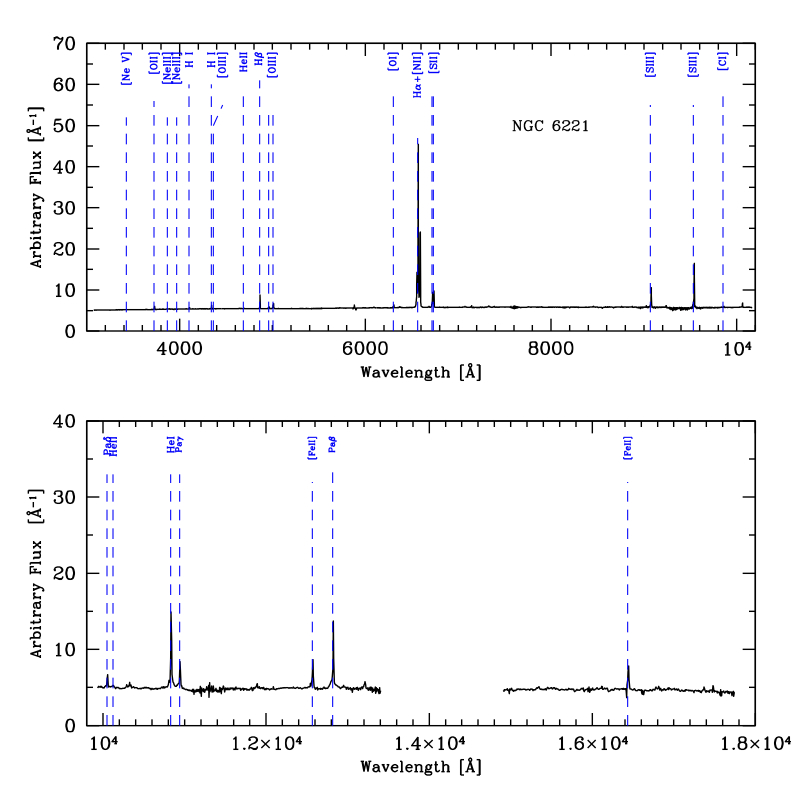}
\end{center}
\textbf{\refstepcounter{figure}\label{fig:1} Figure \arabic{figure}.}{UV, Optical and Near infrared rest frame spectrum of NGC 6221 obtained with
X-Shooter/VLT.  The blue dashed lines show the expected wavelength position of the most relevant emission lines.}
\end{figure}

\begin{figure}
\begin{center}
\includegraphics[width=18.5 cm,angle=0]{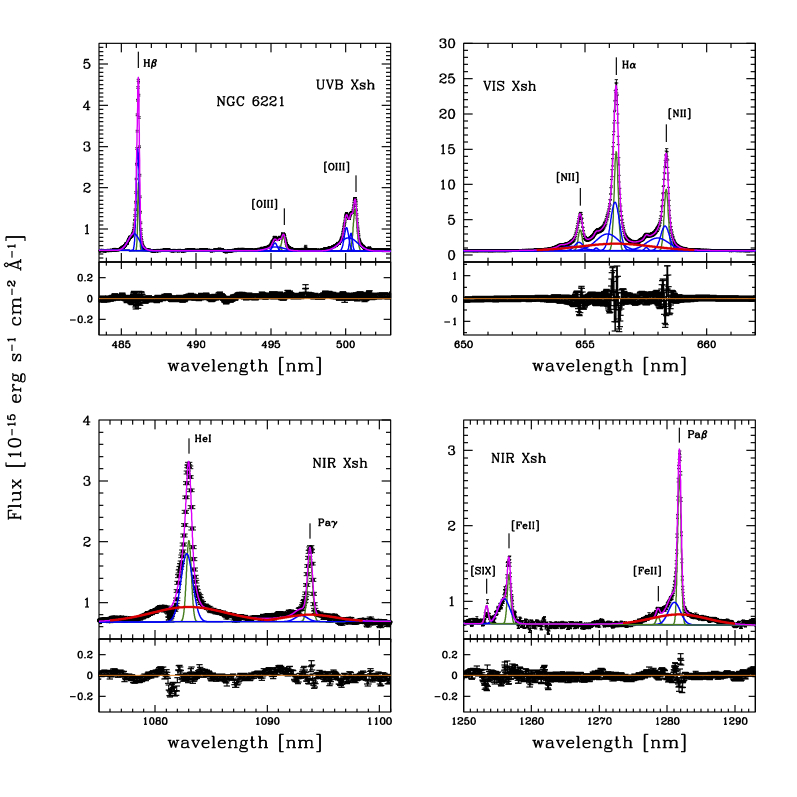}
\end{center}
\textbf{\refstepcounter{figure}\label{fig:2} Figure \arabic{figure}. }{Optical and NIR rest-frame spectra of NGC6221. {\it Top-left}: H$\beta$+[OIII] region.    {\it Top-right}: H$\alpha$+[NII] region.
 {\it Bottom-left}: HeI+Pa$\gamma$ region.    {\it Bottom-right}:  Pa$\beta$+[FeII] region.
The narrow, intermediate and broad components are shown with green, blue and red lines, respectively. The magenta line shows the total fitting model. Vertical lines shows the 
wavelength of the emission lines. Bottom panels show the residuals of the  fitting models.}
\end{figure}

The X-shooter NIR spectrum of NGC 6221 is rich of intense emission lines, both permittßed and forbidden, such as: H$\beta$+[OIII] and H$\alpha$+[NII]
in the optical and HeI (well separated from the Pa$\gamma$) and Pa$\beta$+[FeII]12570 in the NIR (see Figures \ref{fig:1} and \ref{fig:2}).

In the optical region, beside the NLR components, both in the H$\beta$ and H$\alpha$ regions three
intermediate components, having a blueshift up to $\Delta$V=380 km s$^{-1}$, have been found. Moreover,
to significantly better model the  data, a broad H$\alpha$ component (FWHM = 1630 km s$^{-1}$ and  centered with the H$\alpha$ NLR component) was also added (see Figure \ref{fig:2} and Table \ref{tab:1}). Indeed, the F-test gives a probability of 1$\times$10$^{-256}$ that the improvement of the fit obtained including this last broad component is due to statistical fluctuations.  The
H$\alpha$ to H$\beta$ flux ratio of the narrow components confirms the previous estimates of reddening with ${\rm A_V=3}$   (see previous section).

\begin{figure}
\begin{center}
\includegraphics[width=12cm,angle=-90]{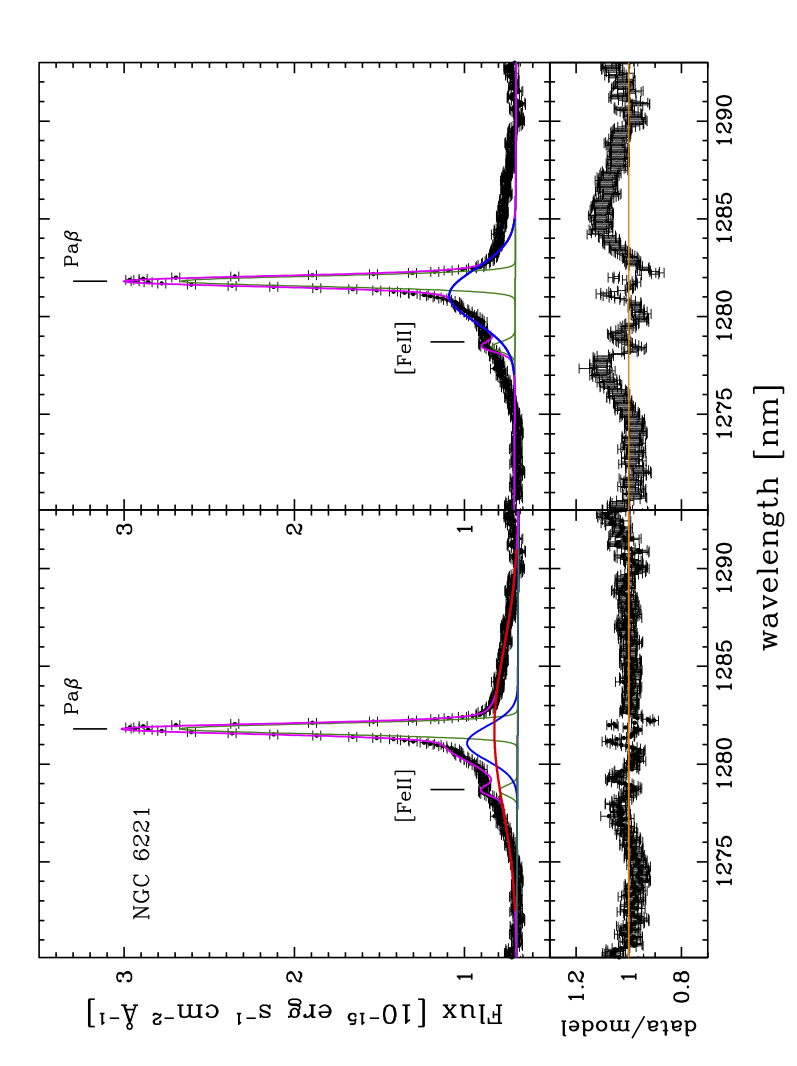}
\end{center}
 \textbf{\refstepcounter{figure}\label{fig:3} Figure \arabic{figure}.}{
 {\it Left}: Best fit of the Pa$\beta$  line of NGC 6221 including a broad (FWHM = 2257 km s$^{-1}$) component. {\it Right}: Same as before without including a broad Pa$\beta$ component.  The narrow, intermediate and broad components are shown with green, blue and red lines, respectively. Lower panels show the data to model ratio.}
\end{figure}

The fitting models of the NIR spectrum are similar to the models found in the optical band. Besides the presence of the NLR components, in each emission line an intermediate component having a blueshift up to $\Delta$V=210 km s$^{-1}$ has been added. Moreover, to significantly improve the modeling of the  data, for each of the HeI, Pa$\gamma$ and Pa$\beta$ lines,  broad ( FWHM = 1430-2260 km s$^{-1}$) components,  centered with the NLR, were also added  (see Figure \ref{fig:2} and Table \ref{tab:1}). As an example, in Figure \ref{fig:3} the fit with, and without, the inclusion of the broad Pa$\beta$ component is shown. The F-test gives a probability of 1$\times$10$^{-118}$ that the improvement of the fit is due to statistical fluctuations.

In summary,  all the lines show narrow components, all sharing the same systemic redshift, with FWHM$<$180 km s$^{-1}$ and some intermediate components showing blueshifts up to $\Delta$V=380 km s$^{-1}$ with respect to the narrow components. This result is in line with the results of \citet{levenson01} and many other authors as described in sect. 2.  Moreover, the H$\alpha$, HeI, Pa$\gamma$ and Pa$\beta$ lines show  significant broad components, with FWHM $\sim$1400-2300 ${\rm km~ s}^{-1}$, having their center in the rest frame measured by the NLR and which can then be attributed to the BLR.  As we are interested in the  detection of these BLR components, we have not carried out either a more detailed kinematic or line fitting analysis of the intermediate and narrow components, which is beyond the scope of this paper.

\section{Discussion and Conclusions}

The detection of faint broad emission line components in AGN2 (or reddened AGN, in general), although difficult, is not uncommon \citep[see e.g.][]{Veilleux97, riffel06,cai10}. As discussed in the introduction, when detected, the width (e.g. the FWHM) of the BLR could be used
to estimate the BH mass of the AGN using single epoch virial relationships. However,  all these relationships use the AGN luminosity as a proxy of the
virial radius (R$\propto$$\sqrt{L}$) but, unfortunately, in AGN2 the luminosity is absorbed by 
surronding material (probably a clumpy torus within few pc from the center; see \citet{burtscher13, marinucci15} and references therein). For these reasons, recently, virial relationships have been calibrated, based on the
use of the very hard X-ray (14-195 keV) luminosity (which is very little affected, if not at all, by absorption) and the measure of the FWHM of the BLR component of the Pa$\beta$ line  which, being emitted in the NIR, is less affected by reddening:
\begin{eqnarray}
{\rm log (M}_{\rm BH}/{\rm M}_\odot)=0.796(\pm 0.031){\rm log}\left[\left( {\frac{\rm FWHM_{Pa\beta}}{10^4\, \rm km\,s^{-1}}} \right)^2    \left(   {\frac{\rm L_{\rm 14-195 keV}}{10^{42}\rm\ erg\,s^{-1}}}  \right)^{0.5}  \right] \nonumber \\
+ 7.611(\pm0.023)~~ (\pm 0.20).
\label{eq:1}
\end{eqnarray}
The resulting observed spread is 0.23 dex, while the intrinsic spread results to be 0.20 dex \citep{lafranca15}.

NGC 6221 has been observed by SWIFT/BAT in the 14-195 keV band and, according to the SWIFT/BAT 70-month catalogue, it has a luminosity log(L$_{14-195}/ {\rm erg~s^{-1}}$) = 42.05 \citep{Baumgartner13}. Therefore, using eq. \ref{eq:1} and our measure of FWHM$_{\rm Pa\beta}$= 2260$^{+100}_{-90}$ km s$^{-1}$, its BH mass results log(M/M$_\odot$)= 6.60 $\pm$0.25. 
As a broad component of the H$\alpha$ has been also measured (FWHM$_{\rm H\alpha}$= 1630$^{+12}_{-11}$ km s$^{-1}$), we can further estimate the BH mass using either a new virial relationship  where the 14-195 keV luminosity and the FWHM of the H$\alpha$ line are instead used (Onori et al. in prep), or the relation by \citet[][their eq. A1 normalized to $f=4.31$]{greene07b}, where both the H$\alpha$ line FWHM and luminosities are used. In the first case a
BH mass of log(M/M$_\odot$) $= 6.4\pm0.3$ is derived. Also
the relation by \citet{greene07b} (once our measure of the luminosity of the broad H$\alpha$ component, ${\rm logL_{H\alpha} = 39.2}$ erg s$^{-1}$, is used) gives a BH mass log(M/M$_\odot$) $= 6.4\pm0.2$. 
All the above reported measures, according to their uncertainties, are in agreement. It should be however  noted that the 0.2 dex lower value  
obtained using the  H$\alpha$ luminosity (if compared to the value obtained using the Pa$\beta$) can be attributed to the effects of the reddening. Indeed, once corrected for reddening, the relation by \citet{greene07b} gives ${\rm log(M/M_\odot) \sim  6.7 }$.

The bolometric luminosity of NGC 6221 has been evaluated through multiwavelength SED fitting  to be ${\rm L_{bol} =  6.2^{+4.3}_{-3.0}\times 10^{42}~erg~s^{-1}}$ \citep{ramosalmeida11}. While, using the 2-10 keV intrinsic luminosity ${\rm L_{2-10} = 6.6\times 10^{41}~erg~s^{-1}}$  (Bianchi et al. in prep) as a proxy, and following the bolometric correction of \citet{marconi04},
${\rm L_{bol} =  1.3\times 10^{43}~erg~s^{-1}}$ is obtained. Therefore, our estimate of a BH mass log(M/M$_\odot$)= 6.6 corresponds to an Eddington ratio
$\lambda_{\rm Edd} = {\rm L_{Bol}/L_{Edd}=0.01-0.03}$.

According to previous studies (see introduction), low luminosity AGN (AGN2 in particular) were expected to populate the low mass regime of the SMBH mass function. Our result of a BH mass smaller than 10$^7$ M$_\odot$ in the starburst/AGN NGC 6221 is in line (together with the results of \citet{lafranca15}) with these studies. 
In this context the virial relation by \citet{lafranca15} which could be used to measure the BH mass in moderately absorbed AGN (mainly AGN2), 
is also useful to derive, in statistically significant  hard X-ray selected samples of AGN2, the BH mass and the Eddington ratio distributions (Onori et al. in prep.), and verify if the galaxy BH mass scale relations are also valid in their case.

\section*{Disclosure/Conflict-of-Interest Statement}

The authors declare that the research was conducted in the absence of any commercial or financial relationships that could be construed as a potential conflict of interest.

\section*{Author Contributions}
The project was originally conceived by Fabio La Franca. All the authors have then contributed to its design, acquisition, interpretation of the data,
and writing of the paper.



\section*{Acknowledgments}
Based on observations made with ESO Telescopes at the La Silla Paranal Observatory under programme ID 093.A-0766.
We thank the anonymous referees for careful reading of the manuscript and numerous helpful suggestions. 

\textit{Funding\textcolon} PRIN/MIUR 2010NHBSBE and PRIN/INAF 2014\_3.

\bibliographystyle{frontiersinSCNS_ENG_HUMS} 
\bibliography{BibFront1111}


\section*{Figures}


\end{document}